 \documentstyle[12pt]{article}
   \newcommand{\be}{\begin{equation}}
   \newcommand{\ee}{\end{equation}}
   \newcommand{\bea}{\begin{eqnarray}}
   \newcommand{\eea}{\end{eqnarray}}

   \def\d{\delta}
   \def\e{\epsilon}
   \def\f{\phi}

   \def\s{\sigma}

   \def\G{\Gamma}

   \def\O{\Omega}

\begin{document}
   \renewcommand{\theequation}{\arabic{equation}}
   \newcommand{\eqn}[1]{eq.(\ref{#1})}
   
   \renewcommand{\section}[1]{\addtocounter{section}{1}
   \vspace{5mm} \par \noindent
     {\bf \thesection . #1}\setcounter{subsection}{0}
     \par
      \vspace{2mm} } %was 5mm
   \newcommand{\sectionsub}[1]{\addtocounter{section}{1}
   \vspace{5mm} \par \noindent
     {\bf \thesection . #1}\setcounter{subsection}{0}\par}
   \renewcommand{\subsection}[1]{\addtocounter{subsection}{1}
   \vspace{2.5mm}\par\noindent {\em \thesubsection . #1}\par
    \vspace{0.5mm} }
   \renewcommand{\thebibliography}[1]{ {\vspace{5mm}\par \noindent{\bf
   References}\par \vspace{2mm}}
   \list
    {\arabic{enumi}.}{\settowidth\labelwidth{[#1]}\leftmargin\labelwidth
    \advance\leftmargin\labelsep\addtolength{\topsep}{-4em}
    \usecounter{enumi}}
    \def\newblock{\hskip .11em plus .33em minus .07em}
    \sloppy\clubpenalty4000\widowpenalty4000
    \sfcode`\.=1000\relax \setlength{\itemsep}{-0.4em} }

   \vspace{4mm}
   \begin{center}
   {\bf D0-D8-F1 in Massive IIA SUGRA} \vspace{1.4cm}

   MARC MASSAR \footnote{massar@tena4.vub.ac.be} 
   and JAN TROOST  \footnote{
  troost@tena4.vub.ac.be;  \,  Aspirant F.W.O.}\\
   {\em Theoretische Natuurkunde, Vrije Universiteit Brussel} \\
   {\em Pleinlaan 2, B-1050 Brussel, Belgium} \\
   \end{center}
   \centerline{ABSTRACT}
   %\end{center}
   \begin{quote}\small
   We present new supersymmetric solutions of massive IIA
   supergravity involving D0-branes, a D8-brane and a string. For the
   bosonic fields we use a  general ansatz with
   SO(8) symmetry.
   \end{quote}
   \section{Introduction}
    Since a few years, D-branes \cite{pol} are an important
   ingredient of string theories. They play a crucial role in the
   more imaginative of recent developments, such as black hole
   entropy counting, M-theory, Matrix-theory and the ADS/CFT
   correspondence. D-branes are defined as spacetime defects on which
   open strings can end. They can also be found as solutions of
   the low-energy effective action of the different string theories,
   i.e. the corresponding supergravity theories.
   
   In \cite{group}, it was pointed out that to find the D8-brane as a
   supergravity solution, you have to turn to Romans' massive IIA supergravity \cite{GP}
   \cite{Romans}, in
   which the mass of the two form in the supergravity multiplet
   simultaneously acts as a cosmological constant. The mass parameter
   is proportional to the dual of the ten form field strength
   associated to the D8-brane.
   
   In general discussions
   of intersecting brane-solutions in supergravity theories,
   the special features of massive IIA were mostly ignored.
   We study some intersections explicitly to see in detail how they arise,
   and specifically, we will concentrate on configurations involving
   D0-branes, D8-branes and a string.

   In section 2 we discuss briefly the relevant massive IIA SUGRA. In
   section 3 we give our ans\"atze and explain our method. In section
   4 we discuss solutions. 

   Recently, a paper which overlaps part\-ly with ours appea\-red in
   the hep-th ar\-chive \cite{bert}.
   
   \setcounter{equation}{0}
   \section{Massive II SUGRA}
   The massive IIA D=10 supergravity \cite{group} \cite{GP}
   \cite{Romans}
    has the same field content as ordinary IIA
   supergravity, namely: a metric g, a dilaton $\sigma$, a  2-form
   $B$, a three-form $C$, a gravitino $\psi$, and a Majorana spinor
   $\lambda$, but the  2-form is massive and contains the degrees of
   freedom of the usual one-form via a generalised Higgs-mechanism.
   Since we consider purely bosonic backgrounds we do
   not write the fermion terms explicitly in what follows.
    As covariant field strengts, we take: 
   \begin{eqnarray}
   G &=& 4 \, dC + 6 M (B)^2  \\
   H &=& 3 \, dB  , 
   \end{eqnarray} 
   where $M$ is the mass parameter.
   The bosonic part of the lagrangian is
   \begin{eqnarray}
    {\cal L} &=& \sqrt{-g} \left( R -\frac{1}{2}(\partial\sigma)^2 -
   \frac{1}{3}e^{-\sigma}H^2 -\frac{1}{12}e^{\frac{1}{2}\sigma}G^2 -
   M^2 e^{\frac{3}{2}\sigma}B^2
   -\frac{1}{2}M^2 e^{{5\over2}\sigma} \right) \nonumber  \\ 
   {} & {} & + \frac{2}{4 ! \, 4 !} \left(  dC dC B + 4 M dC B^3 + \frac{36}{5} M^2
    B ^5\right) . 
   \end{eqnarray}
  where the mass of the two-form also appears as a cosmological constant.
   The procedure to recover ordinary IIA SUGRA (including the one-form) can be
  found in \cite{Romans}. 
   The classical equations of motion of the lagrangian can be put
  in the form: 
   \begin{eqnarray}
   0 &=& -R_{m n} + \frac {1}{16}M^2 e ^{\frac{5 }{2} \s } g_{m
   n}+\frac{1}{2} \partial_m \sigma \, \partial_n \sigma + e^{-\sigma}\left( H_m\,^{r s}
    H_{r s n}- \frac{1}{12} g_{m n} H^2 \right) \nonumber \\
    &&+ 2 M^2 e^{\frac{3}{2}\,\s}\left(B_m\,^r  B_{n r}-\frac{1}{16}g_{m n} B^2 \right) +
   \frac{1}{3}e^{\frac{\sigma}{2}}\left(G_m\,^{r s t}  G_{n r s t }-\frac{3}{32}g_{m n}
   G^2\right), \nonumber \\
   0 &=& - \Box \, \sigma +\frac{5}{4} M^2 e^{ \frac{ 5}{2} \, \s}+\frac{1}{24}
   e^{\frac{\sigma}{2}} G^2-\frac{1}{3}e^{-\sigma} H^2 +\frac{3}{2}M^2 e^{\frac{3}{2}\,
   \sigma}B^2, \nonumber \\
   0 &=& D_r  \left(e^{\frac{\sigma}{2}} G^{r m n p}\right) +\frac{1}{ 4! \, 3} \e^{m n p q r s t u v w}
  G_{q r s t} H_{u v w} , \nonumber \\
   0 &=& D_p  \left( e^{-\sigma}H^{p m n}\right) -M^2 e^{\frac{3}{2} \, \s}
B^{m n} - M
   e^{\frac{\sigma}{2}} G^{ m n p q} B_{p q} \nonumber \\
  {}&{}&+ \frac{1}{4! 4!} \e^{m n p
  q r s t u v w} G_{p q r s} G_{t u v w} . \label{eom}
   \end{eqnarray}
   To impose supersymmetry on
   our solutions we set the supersymmetry
   variations of the fermions to zero -- the variations of the bosonic
fields automatically vanish. The relevant equations are:
   \begin{eqnarray}
  \sqrt{2} \, \delta\lambda &=& - \frac{1}{2} \partial \sigma \cdot \Gamma \e -\frac{5}{8} M
   e^{\frac{5}{4}\,\s} \e +\frac{3}{8} M e^{\frac{3}{4}\,\s} B \cdot \Gamma^{(2)}
   \Gamma_{11} \e 
    + \frac{1}{12} e^{- \frac {\sigma}{2} } H\cdot\Gamma^{(3)}\Gamma_{11}\e
  \nonumber \\
   {}&{}&- \frac{1}{96} e^{\frac{\sigma}{4}} G \cdot \Gamma^{(4)}\e, \nonumber \\
   \delta \psi_m &=& D_m \e -\frac{1}{32} M e^{\frac{5}{4}\,\s} \Gamma_m \e
   -\frac{1}{32}M e^{\frac{3}{4}\,\s}\left( \Gamma_m \, \Gamma^{(2)} \cdot B +14 \Gamma^p 
  B_{p m}\right) \Gamma_{11} \e \nonumber \\
   &&+ \frac{1}{48}e^{-\frac{\sigma}{2}}\left( \Gamma_m \, \Gamma^{(3)} \cdot
   H
   - 9 \Gamma^{p q} H_{p q m} \right) \e \nonumber \\
   &&+ \frac{1}{128} e^{\frac{\sigma}{4}} \left(
   \Gamma_m \Gamma^{(4)} \cdot G + \frac{20}{3}  \Gamma^{p q r} H_{p q r m} \right)
\e.
\label{susyvar}  
 \end{eqnarray}
   \section{Ansatz and method}
   We concentrate on a supersymmetric configuration involving a
   D8-brane and a string perpendicular to it, in the z-direction. We
   impose the following naive projection conditions on the variation
   parameter of supersymmetry:
   \begin{eqnarray}
    \G_z \e &=&-\Gamma_{012345678}  \Gamma_{11} \e = \e 
    \\
   \Gamma_{z0} \Gamma_{11} \e &=&  \e,
   \end{eqnarray}
    These
   conditions break 1/4 of supersymmetry. Taken together they imply
   the projection condition for a D0-brane:
   \begin{eqnarray}
   \G_0 \G_{11} \e &=& \e ,
   \end{eqnarray}
   such that it is possible to add D0-branes without further breaking
   supersymmetry.
  For the bosonic fields we take the most general static ansatz 
  with an $SO(8)$ symmetry and vanishing four-form G:
   \begin{eqnarray}
   ds^2 &=& -k(r,z)^2 dt^2 + l(r,z)^2 dz^2 + m(r,z)^2 (dr^2 + r^2
   d\Omega_7^2 ) \nonumber \\ \sigma &=& \sigma(r,z) \nonumber \\ B
   &=& B_1 \, dz \wedge dt + B_2 \, dt \wedge dr + B_3 \, dz \wedge dr \nonumber \\ 
  M &=&
   \mbox{piecewise constant},
   \end{eqnarray}
    It contains as special cases the  D8-brane
   (where $k(z)=m(z)$),
   the string solution (where $k(r)=l(r)$) and the D0-brane solution (where
   $l(R)=m(R) $ and $R=\sqrt{r^2+z^2})$. The Ramond-Ramond two form field strength
   associated to the D0 brane is contained in the massive B-field,
   such that our ansatz indeed also applies to the single D0-brane
   case. Remark that the mass parameter can take
   different values on either side of a domain wall.
   
   With these
   ans\"atze for the projection conditions and bosonic fields, we
   study the supersymmetry-variations of the fermionic fields.
   Putting these to zero and studying also the integrablity
   conditions on the Killing spinor equation, gives conditions on the
   bosonic fields, enabling us to eliminate all but two of them 
 \footnote{ The technique described differs slightly
   from the one used mostly in the literature, where a more
   restricted ansatz is used, the bosonic equations of motions are
   then solved and the supersymmetry-variations are checked a
   posteriori.}. We choose to keep $\s$ and $\tilde{m} \equiv e^{-\frac{1}{12} \, \sigma} m$:
\begin{eqnarray}
   l&=& e^{\frac{1}{12} \, \s} \tilde{m}^{-2} \\ k&=&
   e^{-\frac{7}{12} \,\s} \tilde{m}^{-4}  \\ m&=& e^{\frac{1}{12}
   \,\s} \tilde{m}  \\ M B_1 &=& -\frac{1}{3} e^{-\frac{4}{3} \, \s }
   \tilde{m}^{-4} (2 \s_z + 15 \frac{\tilde{m}_z}{\tilde{m}})
   \label{B1}  \\ M B_2 &=& \frac{2}{3} e^{-\frac{4}{3} \, \s }
   \tilde{m}^{-4} (\s_r +3 \frac{\tilde{m}_r}{\tilde{m}} )\label{B2}\\
   B_3&=&0
   \\ H_{rzt} &=& 3 \frac{\tilde{m}_r}{\tilde{m}^7}.\label{H}
   \end{eqnarray}

   After a lengthy calculation, many seemingly non-trivial equations of
   motion can then be proven to be linearly dependent. In the end,
   we only have to solve three coupled differential equations. We
   can choose them to be:
   \begin{eqnarray}
   M &=& 3 (\tilde{m}^2)_{,z} \, e^{-\frac{4}{3} \, \sigma}
    \label{meq}
     \\ \frac{\d \cal{L}}{\d \f} &=& 0  \\
    \frac{\d {\cal L} }{ {\d}  g_{rr} } &=& 0,
   \end{eqnarray}
  
    The first equation (\ref{meq}) follows from setting the variation of the
   fermionic fields to zero, while the next two are just two linearly
   independent equations of motion. The explicit expressions for the equations of
motion are:

\begin{eqnarray}
\bigtriangleup_r  (\tilde{m}^6)	 &=&
   -\frac{5}{3} M^2 (\tilde{m} \, e^{\frac{1}{3} \, \sigma})^8  -
   \frac{4}{3} M (\tilde{m} \, e^{\frac{1}{3} \, \sigma})^{10} e^{- 2
   \, \sigma} \sigma_{,z}  \nonumber \\ \bigtriangleup_r  ( e^{- 2 \,
   \sigma}) &=& \frac{20}{3} ( (\sigma_{,r})^2 e^{-2 \, \sigma})
    -  12 (\frac{\tilde{m}_{,r}}{ \tilde{m}})^2 e^{-2 \, \sigma}
    + 16 \frac{\tilde{m}_{,r}}{ \tilde{m}} \sigma_{,r} e^{-2 \, \sigma}
     \nonumber \\
   {} & {} & + \frac{5}{3} M^2 (\tilde{m} e^{\frac{1}{3} \, \sigma})^2
    +  \frac{26}{3} M (\tilde{m} \, e^{\frac{1}{3} \, \sigma})^4 \s_z e^{-2 \s}
     \nonumber \\
   {} &  {} & +\frac{8}{3} (\tilde{m} \, e^{\frac{1}{3} \, \sigma})^6
   e^{-4 \s} (\s_{,z})^2    + 2(\tilde{m} \,
   e^{\frac{1}{3} \, \sigma})^6 e^{-4 \s} \s_{,zz}. \label{difeq}
   \end{eqnarray}
Any solution of
   these coupled differential equations is a solution of
   massive IIA SUGRA preserving (at least) 1/4 supersymmetry.

   \section{Solutions}
   Although the differential equations are difficult to solve in
   general, in specific cases, they simplify drastically. 
  Our analysis covers a lot of cases already
   discussed in the literature \cite{group} \cite{bert} 
   \cite{Lu} \cite{Ruiz}, and contains  new solutions.
   \subsection{$ H=0 $}
   Solving the equation $H=0$ (\ref{H}), and
    equation (\ref{meq}) demonstrates that only the
   flat D8-brane and the isolated D0-brane are solutions.
   You do not find, for
   instance, an isolated D0-brane in a D8-brane background, which you
   might expect to have the required symmetry. But that configuration
   would be inconsistent, as explained in \cite{polstrom}.
   \subsection{ $B_2 = 0$}
   Using the $B_2$ equation (\ref{B2}), and the mass e\-qua\-tion
   (\ref{meq}), and then ana\-ly\-sing the equations of motion, we
   find that we are left with the following simple differential
   equation:
   \begin{eqnarray}
   e^{ -2 \, \s }  &=& h (r) + M z \nonumber \\
    \bigtriangleup_r (h (r)) &=& - M^2.
   \end{eqnarray}
   Adding in the appropriate source term, we find the following
   solution:
   \begin{eqnarray}
   ds^2  &=& - (M z + 1+ \frac{k_f}{r^6} -\frac{ M^2 r^2}{16})^{-\frac{3}{4} } (dt^2+dz^2)
  \nonumber\\ {}&{}&
          + (M z + 1+ \frac{k_f}{r^6} - \frac{M^2 r^2}{16})^{\frac{1}{4} } (dr^2 + r^2 d \O_7^2)
   \nonumber \\
    e^{-2 \, \s} &=& 1+ \frac{k_f}{r^6} - \frac{M^2 r^2}{16} + M z \nonumber \\
     B &=& -\frac{1}{2} (M z + 1+ \frac{k_f}{r^6} - \frac{M^2 r^2}{16})^{-1} dz \wedge dt \nonumber \\
   \end{eqnarray}
    Independently, the authors of \cite{bert} found this solution
 in a different approach. It is clear that with the
   source terms we chose, in the massless case we recover the
   ordinary string solution \cite{Ruiz}. But the interpretation of the solution
   with the mass term is not  clear. Firstly, we do not
  see an interpretation for the assymptotic geometry. Secondly, the metric
   components of the z and t coordinate are equal, leading the authors of
   \cite{bert} to name this solution a massive string solution, but there is
   no obvious $SO(1,1)$ symmetry, since the metric components
    depend on the z-coordinate. 
   \subsection{$B_1=0$}
    An analogous analysis gives:
   \begin{eqnarray}
   \tilde{m} &=& n (r) e^{-\frac{2}{15} \, \s}\nonumber \\
     e^{ -\frac{8}{5} \, \sigma } &=&  2 M z \, n^{-2} \nonumber \\
   \bigtriangleup_r (n^5 ) &=& 0
   \end{eqnarray}
   Taking appropriate source terms, and making a simple z--coordinate
transformation, we find the solution:
   \begin{eqnarray}
   ds^2 &=& - ( M z )^{\frac{1}{8}}
   (1+\frac{k_f}{r^6})^{-\frac{13}{8}} dt^2 
          + ( M z )^{\frac{9}{8}}  (1+\frac{k_f}{r^6})^{-\frac{5}{8}} dz^2 \nonumber \\
  & &
          +  ( M z )^{\frac{1}{8}} (1+\frac{k_f}{r^6})^{\frac{3}{8}} (dr^2 + r^2 d \O_7^2)
   \nonumber \\ 
   e^{\frac{4}{5} \, \s} &=& (M z)^{-1}(1+\frac{k_f}{r^6})^{\frac{1}{5}}  \nonumber \\
     B &=& -\frac{3 }{M} \frac{k_f r^5  M z}{(k_f+r^6)^2} dt \wedge dr \nonumber \\
     {}&={}& -\frac{1}{2}(1+\frac{k_f}{r^6})^{-1} dz \wedge dt
  +\frac{1}{2M}d\left(\frac{Mz}{1+\frac{k_f}{r^6}}dt\right)
  \end{eqnarray}
    The
   solution assymptotically goes to a flat D8-brane \cite{group}. The other
   building block in the solution is a harmonic superposition of
   D0-branes and a string. Note that it does not follow from the ordinary harmonic 
  superposition rules \cite{harm}. The two-form B is split  in the last line
in a part that reminds of the string solution and an exact
part that can be interpreted as the D0-brane gauge field $C_{(1)}$.
   \subsection{ $M=0$ }
   When $M=0$,  we get ordinary IIA SUGRA \cite{GP} \cite{Romans}.
    Little analysis yields:
   \begin{eqnarray}
     \tilde{m} & \equiv &  n (r) ^{1/6}  \nonumber \\
    \bigtriangleup_r (h) +   n \, h_{,zz} &=& 0 \nonumber \\
   \bigtriangleup_r (n) &=& 0        \nonumber \\ C_{(1)}
   &=&  h^{-1} dt
   \end{eqnarray}
   where $ e^{-2 \, \s}  \equiv  n \, h^{-\frac{3}{2}} $.
  This set of differential equations incorporates
   the flat string solution ($h=1$ and n harmonic), the lonely D0-brane solution
   ($n=1$ and h harmonic), as well as the harmonic superposition of both
   ($h=n$). We want to indicate one more possibility,
   namely the following solution:
   \begin{eqnarray}
   ds^2 &=& - (a |z| )^{-\frac{7}{8}}
   (1+\frac{k_f}{r^6})^{-\frac{3}{4}} dt^2
          +(a |z| )^{\frac{1}{8}}  (1+\frac{k_f}{r^6})^{-\frac{3}{4}} dz^2 \nonumber \\
  &  &
          + (a |z| )^{\frac{1}{8}} (1+\frac{k_f}{r^6})^{\frac{1}{4}} (dr^2 + r^2 d \O_7^2)
    \nonumber \\
    e^{-2 \, \s} &=&
   (1+\frac{k_f}{r^6}) ( a |z|)^{-\frac{3}{2}} \nonumber
   \\
    M &=& 0 \nonumber \\
    B &=& -\frac{1}{2} (1+\frac{k_f}{r^6})^{-1} dz \wedge dt \nonumber \\
    C_{(1)} &=& \frac{1}{a | z |}   dz 
   \end{eqnarray}
   This solution consists of an eight-dimensional wall of D0-branes and a
string perpendicular to it. The behavior of the potential is similar
to that of a point charge in one dimension.
   \section{Conclusions}
   We considered the problem of finding new supersymmetric SUGRA
   solutions preserving 1/4 supersymmetry. Starting from a quite general
   ansatz, corresponding to a  superposition of D0-branes,  a string
   and a D8-brane with $SO(8)$ symmetry, we studied the supersymmetry
variations of the fermionic fields and thus simplified the equations
   of motion. We cover all cases of combinations of these branes in
   the literature. Moreover, we discovered some new solutions. One of them
 reduces in the massless case to the ordinary
fundamental 
string solution, while the interpretation in the massive case is unclear.
This solution was independently found in \cite{bert}. Another solution
consists of a harmonic superposition of D0-branes and a string in a D8-brane
background.
 Our system of differential equations  may well contain other
   non-trivial configurations with interesting physical
   interpretations. For instance the phenomenon of the creation of a
   string in the D0-D8 brane system \cite{crea} could be approached
   with our ansatz.  Moreover it would be interesting to apply the
   masssive T-duality rules \cite{group} \cite{BHO}  to our new
   solutions to gain more insight in their interpretation. 
We also plan to study other supergravity configurations with our approach.
\noindent
   
   \vspace{1cm}
   
   \noindent {\bf Acknowledgments}: We would like to thank Marco
   Billo, Ben Craps, Frederik Denef, Frederik Roose, Rodolfo Russo
   and Walter Troost for  discussions. We gladly made use of the
   GRTensorM 1.2 Mathematica package \cite{MPL}. This work was supported in part by the
European Commission TMR programme ERBFMRX-CT96-0045 in which the authors are
associated to K.U. Leuven.


\newpage
   \begin{thebibliography}{99}
   \bibitem{pol} J. Polchinski, Phys. Rev. Lett. {\bf 75}, 4724 (1995) hep-th/9510017;
    J. Polchinski, hep-th/9611050
    \bibitem{group} E. Bergshoeff, M. de Roo, M. B. Green, G. Papadopoulos and
   P.K. Townsend, Nucl. Phys {\bf B470} (1996) 113, hep-th/9601150.
   \bibitem{GP} F. Giani and M. Pernici, Phys. Rev {\bf D30} (1984) 325
   \bibitem{Romans} L. J. Romans, Phys. Lett. {\bf B169} (1986) 374.
   \bibitem{bert} B. Janssen, P. Meessen and T. Ortin, hep-th/9901078
   \bibitem{Killing} H. L\"u, C. N. Pope and J. Rahmfeld, hep-th/9805151
   \bibitem{harm} G. Papadopoulos and P. Townsend, Phys. Lett. {\bf B380}
   (1996) 273, hep-th/9603087; A. Tseytlin, Nucl. Phys. {\bf B475}
   (1996) 179, hep-th/9604035; J. Gauntlett, D. Kastor and J.
   Traschen, Nucl. Phys {\bf B478} (1996) 544, hep-th/9604179
   \bibitem{Lu} M.J. Duff and J. X. Lu, Nucl. Phys. {\bf B411} (1994) 473,
   hep-th/9306052 and references
   therein.
   \bibitem{polstrom} J. Polchinski and A. Strominger, Phys. Lett. {\bf 388}
  (1996) 736  hep-th/9510227;
                      A. Strominger, Phys. Lett. {\bf B383} (1996) 44,
   hep-th/9512059
   \bibitem{Ruiz} A. Dabholkar, G. W. Gibbons, J. A. Harvey and F. Ruiz-Ruiz,
   Nucl. Phys. {\bf B340} (1990) 33.
    \bibitem{crea} C. Bachas, M.Douglas, M. Green, {\bf JHEP} 9707(1997) 002,
  hep-th/9705074;Pei-Ming Ho
 and Yong-Shi Wu,Phys.Lett. {\bf B420} (1998) 43, hep-th/9708137; Pei-Ming Ho, Miao Li and Yong-Shi
 Wu, Nucl.Phys. {\bf B525} (1998) 146, hep-th/9706073; 
   U.Danielsson, G Ferreti, I. Klebanov, Phys. Rev. Lett. {\bf 79}
  (1997) 1984, hep-th/9705084; O. Bergman, M. Gaberdiel, G. Lifschytz,
  Nucl. Phys. {\bf B 509} (1998) 194,
   hep-th/9705130;  M. Billo, P. Di Vecchia, M. Frau, A. Lerda, I. Pesando,
 R. Russo, S. Sciuto, Nucl.Phys. {\bf B526} (1998) 199, hep-th/9802088
   \bibitem{BHO} E. Bergshoef, C. Hull and T. Ort\'{\i}n,
   Nucl. Phys {\bf B451} (1995) 547, hep-th/9504081 
   \bibitem{MPL} P. Musgrave, D. Pollney and K. Lake, GRTensor II,
http://www.astro.queensu.ca /~grtensor/
   \end{thebibliography}
   \end{document}